\documentclass[twocolumn,pra,unsortedaddress,superscriptaddress,floatfix,citeautoscript,nofootinbib,longbibliography]{revtex4-1}
\usepackage{adjustbox}
\usepackage{dcolumn}
\usepackage{latexsym,epsfig,graphicx,dcolumn,subfigure,comment,ulem}
\usepackage{amsmath,eqnarray,amssymb,amsbsy}
\usepackage{xcolor,color}
\usepackage[colorlinks,urlcolor=blue,citecolor=blue]{hyperref}

\begin{document}

\title{Anomalous Fraunhofer-like patterns in  quantum anomalous Hall Josephson junctions}

\author{Junjie Qi}
\email{qijj@baqis.ac.cn}
\affiliation{Beijing Academy of Quantum Information Sciences, 
	Beijing 100193, China}
 \author{Haiwen Liu}
 \affiliation{Center for Advanced Quantum Studies, Department of Physics, Beijing Normal University, Beijing 100875,
China}
\author{Jie Liu}
\affiliation{School of Physics, Xi’an Jiaotong University, Ministry of Education Key Laboratory for Non-Equilibrium Synthesis
and Modulation of Condensed Matter, Xi’an 710049, China}
\author{Hua Jiang}
\affiliation{
Institute for Nanoelectronic Devices and Quantum Computing, Fudan University, Shanghai 200433, China}
\author{Dong E. Liu}
\affiliation{State Key Laboratory of Low Dimensional Quantum Physics, Department of Physics, Tsinghua University, Beijing 100084, China}
\affiliation{Beijing Academy of Quantum Information Sciences, 
	Beijing 100193, China}
\affiliation{Frontier Science Center for Quantum Information, Beijing 100084, China}
\affiliation{Hefei National Laboratory, Hefei 230088, China}
 \author{Chui-Zhen Chen}
 \email{czchen@suda.edu.cn}
 \affiliation{School of Physical Science and Technology, Soochow University, Suzhou 215006, China}
\affiliation{
Institute for Advanced Study, Soochow University, Suzhou 215006, China}
\author{Ke He}
\affiliation{State Key Laboratory of Low Dimensional Quantum Physics, Department of Physics, Tsinghua University, Beijing 100084, China}
\affiliation{Beijing Academy of Quantum Information Sciences, 
	Beijing 100193, China}
\affiliation{Frontier Science Center for Quantum Information, Beijing 100084, China}
\affiliation{Hefei National Laboratory, Hefei 230088, China}
\author{X. C. Xie}
\affiliation{  }
\affiliation{
Institute for Nanoelectronic Devices and Quantum Computing, Fudan University, Shanghai 200433, China}
\affiliation{Hefei National Laboratory, Hefei 230088, China}

\begin{abstract}

The intriguing interplay between topology and superconductivity has attracted significant attention, given its potential for realizing topological superconductivity.  In the quantum anomalous Hall insulators (QAHIs)-based junction, the supercurrents are carried by the chiral edge states, characterized by a $2\Phi_0$ magnetic flux periodicity  ($\Phi_0 = h/2e$ is the flux quantum, $h$ the Planck constant, and $e$ the electron charge). However, experimental observations indicate the presence of bulk carriers in QAHI samples due to magnetic dopants.  In this study, we reveal a systematic transition from edge-state to bulk-state dominant supercurrents as the chemical potential varies from the bulk gap to the conduction band. This results in an evolution from a $2\Phi_0$-periodic oscillation pattern to an asymmetric Fraunhofer pattern. Furthermore, a novel Fraunhoher-like pattern emerges due to the coexistence of chiral edge states and bulk states caused by magnetic {\color{black}domains}, even when the chemical potential resides within the gap. These findings not only advance the theoretical understanding but also pave the way for the experimental discovery of the chiral Josephson effect based on QAHI doped with magnetic impurities.

\end{abstract}

\maketitle

\section{Introduction}\label{sec:introduction}

 In recent decades, research on quantum anomalous Hall insulators (QAHIs) has flourished, driven not only by the intriguing physical phenomena they exhibit but also by their potential for various technological applications \cite{Haldane,QAH2,QAH3,XDH}. {\color{black}These applications include serving as platforms for realizing topological quantum computation, resistance standards, and dissipation-free interconnects} \cite{QAH3}.
  In a QAHI,  the bulk behaves as an insulator, while electrons can propagate along the edges of the sample. {\color{black} These edge states, termed chiral edge states, propagate unidirectionally along the edge.}
  Consequently, QAHI is distinguished by a quantized Hall effect, even in the absence of external magnetic fields. The existence of QAHI states has been theoretically predicted in a range of materials \cite{QAHt1,QAHt2,QAHt3} and experimentally verified in Cr-doped topological insulator films for the first time \cite{QAH1}. Subsequently, QAHI has been experimentally demonstrated in various materials, including 
Cr-doped $\rm{(Bi,Sb)_2Te_3}$ films \cite{QAH11,QAH12,QAH13,QAH14,QAH15,QAH16,QAH17,QAH18,QAH19}, V-doped  $\rm{(Bi,Sb)_2Te_3}$ films \cite{QAH4,QAH12,QAH21,QAH22,QAH23,QAH24}, $\rm{MnBi_2Te_4}$ \cite{QAH5,QAH31,QAH32}, and recently moiré superlattice systems \cite{QAH6,QAH7,QAH8}. 
Remarkably, the proximity of a QAHI to an s-wave superconductor (SC)  was predicted to potentially realize a chiral topological superconductor within these heterostructures via the superconducting proximity effect \cite{TSC7,CMF1}.
Therefore, QAHI/SC heterostructures have garnered considerable attention as promising platforms for realizing topological superconductors \cite{E1,E2,E3,TSC1,TSC2,TSC3,TSC4,TSC5,TSC6,TSC7,YNX,Tanaka,ZYF}.
More interestingly, when a QAHI is sandwiched by two QAHI/SC heterostructures, {\color{black}supercurrent is carried by the chiral edge states, termed a chiral Josephson junction (JJ) in our paper. }
 Investigating JJs based on QAHIs provides insights into the interplay between topology and superconductivity, offering the potential for novel phenomena and applications in quantum information processing and topological quantum computing \cite{Asano,YQ,Ryota,QC}.

\begin{figure}[ht!]
\centering
\includegraphics[width=1\linewidth]{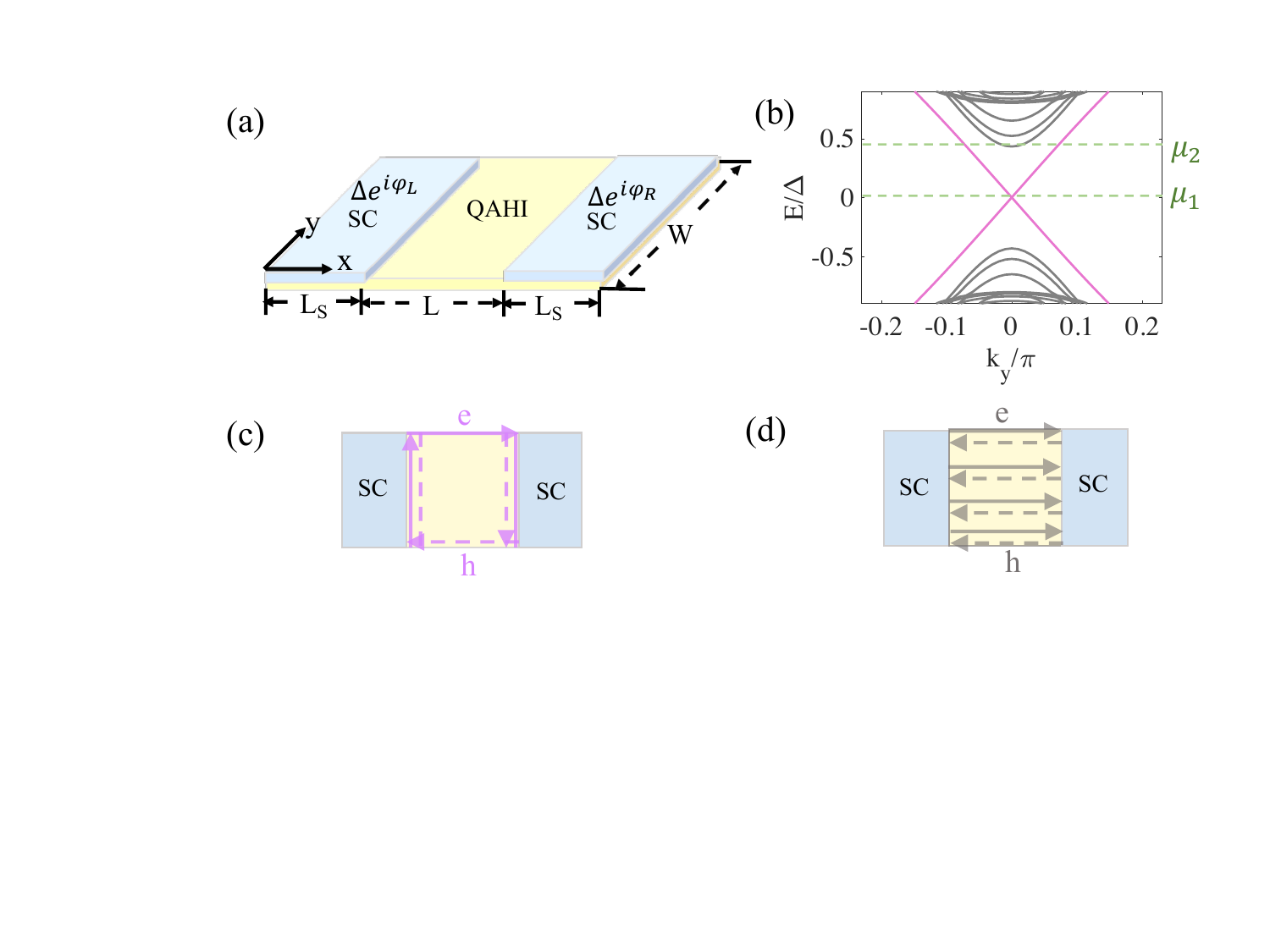}
\caption{(a) Schematic diagram of a QAHI-based JJ. The QAHI is partially covered by two superconducting electrodes, both composed of QAHI/SC heterostructures. The central region and the superconducting region have sizes of $ W \times L$ and $ W \times L_s$, respectively.  (b) The QAHI energy spectrum with a length of $L=40a$ is calculated by using a lattice model described by Eq.~\ref{eq:2}. $\mu_1=0$ and $\mu_2=0.42\Delta$ are the chemical potentials, indicated by the green dashed lines.  (c) Schematic diagram of CARs in real space, where Andreev pairs are constituted by electrons and holes propagating along opposing edges.    (d) Schematic diagram of  LARs in real space, where Andreev pairs formed by bulk carriers exhibit uniform distribution within the central region. In Figs. (b)-(d), we use purple (gray) lines to represent edge (bulk) carriers, respectively.}
\label{fig:1}
\end{figure}

 Generally, the supercurrent in JJs  is generated by the Andreev reflection process, which involves the conversion of an electron into a hole at the superconductor interface \cite{AR}. This process is accompanied by the creation of a Cooper pair in the superconducting region. {\color{black} As shown in Fig.~\ref{fig:1}(d), local Andreev reflections (LARs) occur uniformly in a conventional JJ.
 Thus, the critical Josephson current $I_c$, which is carried by bulk carriers, oscillates with the external magnetic flux $\Phi$ due to the interference of the supercurrents flowing through the junction in the presence of the magnetic flux. This oscillatory behavior gives rise to a quantum interference pattern known as the Fraunhofer pattern. The Fraunhofer pattern is characterized by a central lobe with a width of $2\Phi_0$ and decaying side lobes. Here, $\Phi_0 = h/2e$ is the flux quantum, $h$ the Planck constant, and $e$ the electron charge.  When time-reversal symmetry is preserved, the Fraunhofer pattern is symmetric, meaning $I_c(\Phi) = I_c(-\Phi)$. }
  
   On the other hand, in a chiral JJ depicted  in Fig.~\ref{fig:1}(a), {\color{black} the presence of gapped bulk states inhibits  the bulk supercurrent. However, supercurrents can propagate through the chiral edge states even when the chemical potential resides within the bulk gap.} Due to the chiral nature of edge states,  electrons on the top edge are transformed into  holes propagating along the bottom edge (see Fig.~\ref{fig:1}(c)).  This conversion of electrons and holes takes place in the spatially separated edges, known as crossed Andreev reflections (CAR) {\color{black}\cite{CAR1,CAR2,f2,f6,f8}}. {\color{black}  This process is reminiscent of the Aharonov-Bohm (AB) oscillations \cite{AB1}, where electrons traveling along different paths enclose a magnetic flux $\Phi$, causing resistance to oscillate with the magnetic flux exhibiting a periodicity of $2\Phi_0$ \cite{AB2}.} Consequently, the critical Josephson current $I_c$ in response to the external magnetic flux $\Phi$ can be described theoretically by the periodicity of $2\Phi_0$ \cite{f1,f2,f3,f4,f5,f6,f7,f8}. 
 On the other hand, experimental observations have confirmed the importance of the bulk states, induced by magnetic dopants, on the transport properties of QAHIs \cite{QAH2,QAH3}. However, previous studies did not consider the effects of these bulk states on the quantum interference patterns in chiral JJs. Therefore, research addressing this question is highly desirable.
 
In this work, we systematically explore a JJ based on a QAHI with magnetic dopants, aiming to address the above question. The Josephson current $I_s$ is calculated by the recursive Green-function method. First,  we present the critical Josephson current $I_c$ as a function of the magnetic flux $\Phi$ for various chemical potentials $\mu$. We find an evolution of quantum interference patterns from  a $2\Phi_0$-periodic oscillation pattern to an asymmetric Fraunhofer pattern. The two typical interference patterns result from different Andreev reflection processes as shown in Figs.~\ref{fig:1}(c)-(d). Second, we simulate the effects of {\color{black}magnetic domains} on the two interference patterns. We find that the asymmetric Fraunhofer pattern is very robust against the {\color{black}domains}, while the $2\Phi_0$-periodic oscillation pattern is susceptible to modest magnetic fluctuations.  
Remarkably, a novel Fraunhofer-like pattern, with periods twice those of the conventional one, emerges when the chiral edge state and magnetic {\color{black}domain} states coexist at $\mu=0$. 
The observed Fraunhofer-like pattern is expected to extend beyond our model and could be prevalent in materials doped with magnetic impurities, suggesting potential for further experimental verification.

The paper is organized as follows. We introduce the model and methods employed in this work in Sec.~\ref{sec:model}. Sec.~\ref{sec:results} 
presents the key findings of our calculations and the corresponding remarks. 
More specifically,  Sec.~\ref{subsec:B} discusses two distinct quantum interference patterns. The effects of {\color{black}magnetic domains}, which are covered in 
Sec.~\ref{subsec:C},  inevitably occur in the magnetically doped topological insulators. We find that the $2\Phi_0$-periodic oscillation effect is destroyed by modest magnetic fluctuations. 
A novel Fraunhofer-like pattern is observed even when $\mu=0$.
Sec.~\ref{sec:conclusion} is the conclusion.

\section{Model and methods}\label{sec:model}

As depicted in Fig.~\ref{fig:1}(a), the QAHI-based junction comprises three parts: two superconducting electrodes and a central region with QAHI.
The $4 \times 4$ low-energy effective Hamiltonian describing magnetic topological insulator thin films in the central region is given by \cite{QAHt3, WJ}
\begin{align}\label{eq:1}
	H(\mathbf{k}) =\hbar v_F\left(k_y \sigma_x \tau_z-k_x \sigma_y \tau_z\right)+m(\mathbf{k}) \tau_x+M_z\sigma_z
\end{align}
in the basis of $\psi_\mathbf{k}=\left[\psi_{\mathbf{k},t\uparrow}, \psi_{\mathbf{k},t\downarrow}, \psi_{\mathbf{k}, b \uparrow}, \psi_{\mathbf{k}, b \downarrow}\right]^{\mathrm{T}}$ with the wave vector $\mathbf{k}$.  $\uparrow$ ($\downarrow$) represents the spin direction, and $t$ ($b$) denotes  the top (bottom) layer. The symbols $\sigma_{x,y,z}$ and $\tau_{x,y,z}$ correspond to the Pauli matrices for spin and layer, respectively.
$M_z$  represents the exchange field along the $z$ axis induced by the ferromagnetic ordering. 
The term $m(\mathbf{k})=m_0-m_1 \mathbf{k}^2$ describes the coupling between the top and bottom {\color{black} layers}.
The system is in the QAHI phase with the Chern number $C=\mathrm{sgn}(M_z)$ when $\left|M_z\right|>\left|m_0\right|$, while it is a normal insulator with $C=0$ when $\left|M_z\right|<\left|m_0\right|$.  

In our numerical simulations, we discrete the Hamiltonian in Eq.~\ref{eq:1} to a tight-binding model in a square lattice as \cite{YQ}
\begin{align}\label{eq:2}
	H&=\sum_{ij}\left(\psi_{i}^{\dagger} t_0 \psi_{i}+\psi_{i}^{\dagger} t_x \psi_{i+\hat{x}}+\psi_{i}^{\dagger} t_y \psi_{i+\hat{y}}+\mathrm{H.c.} \right)
\end{align}
where $\hat{x}$ ($\hat{y}$) the unit vector along the $x$ ($y$) axis and $i$ the site index. The 4$\times$4 matrices $t_{0,x,y}$ are given by
\begin{equation}\label{eq:3}
	\begin{aligned}
		t_0 & =\left(m_0-4 \frac{m_1}{a^2}\right) \tau_x \sigma_0+M_z\tau_0\sigma_z, \\ 	
		t_x & =\left(\frac{m_1}{a^2} \tau_x \sigma_0+\frac{\mathrm{i} v_{\mathrm{F}}}{2 a} \tau_z \sigma_y\right)e^{i\phi_{i,i+\hat{x}}}, \\ 
		t_y & =\frac{m_1}{a^2} \tau_x \sigma_0-\frac{\mathrm{i} v_{\mathrm{F}}}{2 a} \tau_z \sigma_x .
	\end{aligned}
\end{equation}
where $t_x$ gains a phase $\phi_{ij}= \pi \int_{i}^{j}\mathbf{A}\cdot d\mathbf{l}/\Phi_{0}$  due to the magnetic field $B$ in z direction with the vector potential $\mathbf{A}=(-yB,0,0)$,
and $\Phi_{0}=h/2e$ is the flux quantum  with the Planck constant $h$  and the elementary charge $e$. Here $\sigma_0$ ($\tau_0$) denotes the $2 \times 2$ identity matrix. The total magnetic flux of the central region is denoted as $\Phi$. {\color{black} 
Utilizing Eqs.~\ref{eq:2} and ~\ref{eq:3}, we can visualize the energy spectrum in the QAHI when $\Phi=0$, depicted in Fig.~\ref{fig:1}(b).  The bulk states (gray) are gapped, accompanied by the gapless edge states (purple).}

In proximity to an s-wave SC,  a finite pair amplitude is induced in the QAHI.  The Bogoliubov-de Gennes (BdG) Hamiltonian of QAHI/SC heterostructures in Nambu space is given by  $   H_{\mathrm{SC}}=\allowbreak  \sum_{\mathbf{k}} \Psi_{\mathbf{k}}^{\dagger} H_{\mathrm{SC}}(\mathbf{k}) \Psi_{\mathbf{k}}/2
$ with $\Psi_\mathbf{k}=[(\Psi_{\mathbf{k},t\uparrow}, \Psi_{\mathbf{k},t\downarrow}, \Psi_{\mathbf{k}, b \uparrow}, \allowbreak \Psi_{\mathbf{k}, b \downarrow}),(\Psi_{\mathbf{-k},t\uparrow}^{\dagger}, \Psi_{\mathbf{-k},t\downarrow}^{\dagger}, \Psi_{\mathbf{-k}, b \uparrow}^{\dagger}, \Psi_{\mathbf{-k}, b \downarrow}^{\dagger})]^{\mathrm{T}}$
and 
\begin{center}
	\begin{equation}\label{eq:4}
		\begin{aligned}
			H_{\mathrm{SC}} (\mathbf{k})& =\left(\begin{array}{cc}
				H(\mathbf{k})-\mu_s & \Delta \\
				\Delta^{\dagger} & -H^*(-\mathbf{k})+\mu_s
			\end{array}\right), \\
			\Delta & =\left(\begin{array}{cc}
				i e^{i\varphi_{L(R)}}\Delta_t \sigma_y & 0 \\
				0 &ie^{i\varphi_{L(R)}}\Delta_b  \sigma_y
			\end{array}\right) .
		\end{aligned}
	\end{equation}
\end{center}

Here, $\mu_s$ is the chemical potential of the superconduting regions, $\varphi_L$ and $\varphi_R$ are the superconducting phases in the left and right sides of the junction. The physical properties only depend on  the superconducting phase difference $\varphi=\varphi_L-\varphi_R$. $\Delta_t$ and $\Delta_b$ denote the induced superconducting pairing potentials on the top and bottom {\color{black} layers}, respectively.

The QAHI regions in Nambu space can be represented as 

	\begin{equation}\label{eq:5}
			H_{\mathrm{BdG}} (\mathbf{k}) =\left(\begin{array}{cc}
				H(\mathbf{k})-\mu & 0 \\
				0 & -H^*(-\mathbf{k})+\mu
			\end{array}\right)
	\end{equation}

Here, $\mu$ is the chemical potential of the QAHI region. For simplicity, the parameters are set as follows: lattice constant $a=1$, Fermi velocity $\upsilon_F=1$, $\Delta_t=0.5$ and $\Delta_b=0$, $m_1=2\Delta_t a^2$, and $m_0=-0.2\Delta_t$ in this work  and  both the QAHI and superconducting regions have dimensions $W \times L(L_s) = 100a \times 40a$.  In the following calculations, we use $\Delta$ instead of $\Delta_t$.

The Josephson current $I$ in the QAHI region can be calculated by the recursive Green-function method and expressed as  \cite{M1,M2,M3}

\begin{equation}\label{eq:6}
I=-\frac{ieK_B T}{\hbar} \sum_{{\color{black}n}} \operatorname{Tr}\left[\hat{T}_{x} \hat{G}_{x+1,x}(i\omega_n) - \hat{T}^{\dagger}_{x} \hat{G}_{x,x+1}(i\omega_n)\right]
\end{equation}

where $e$ is the electron charge, $k_B$ is the Boltzmann constant,  
 $\hbar$ is the {\color{black}reduced} Planck constant, and $T$ is the temperature. The hopping term $\hat{T}_x$ in Nambu space  is given by $\hat{T}_x=\left(\begin{array}{cc}
t_x & \\
& -t_x^{*}
\end{array}\right)$, where $t_x$ originates from Eq.~\ref{eq:3}. The Green's function is expressed as $\hat{G}(i\omega_n)=\left(i \omega_n-H_{B d G}-\Sigma_L-\Sigma_R\right)^{-1}$ with the Matsubara frequency $\omega_n=(2 n+1) \pi k_B T$.  {\color{black} The index 
$n$ in the sum represents a summation over Matsubara frequencies, where 
$n$ spans all integer values.}
Here, $\Sigma_{L(R)}$ denotes the self-energy induced by the coupling between the left (right) superconducting leads and the QAHI region, which can be calculated numerically \cite{lead}.

\begin{figure}[ht!]
\centering
\includegraphics[width=1\linewidth]{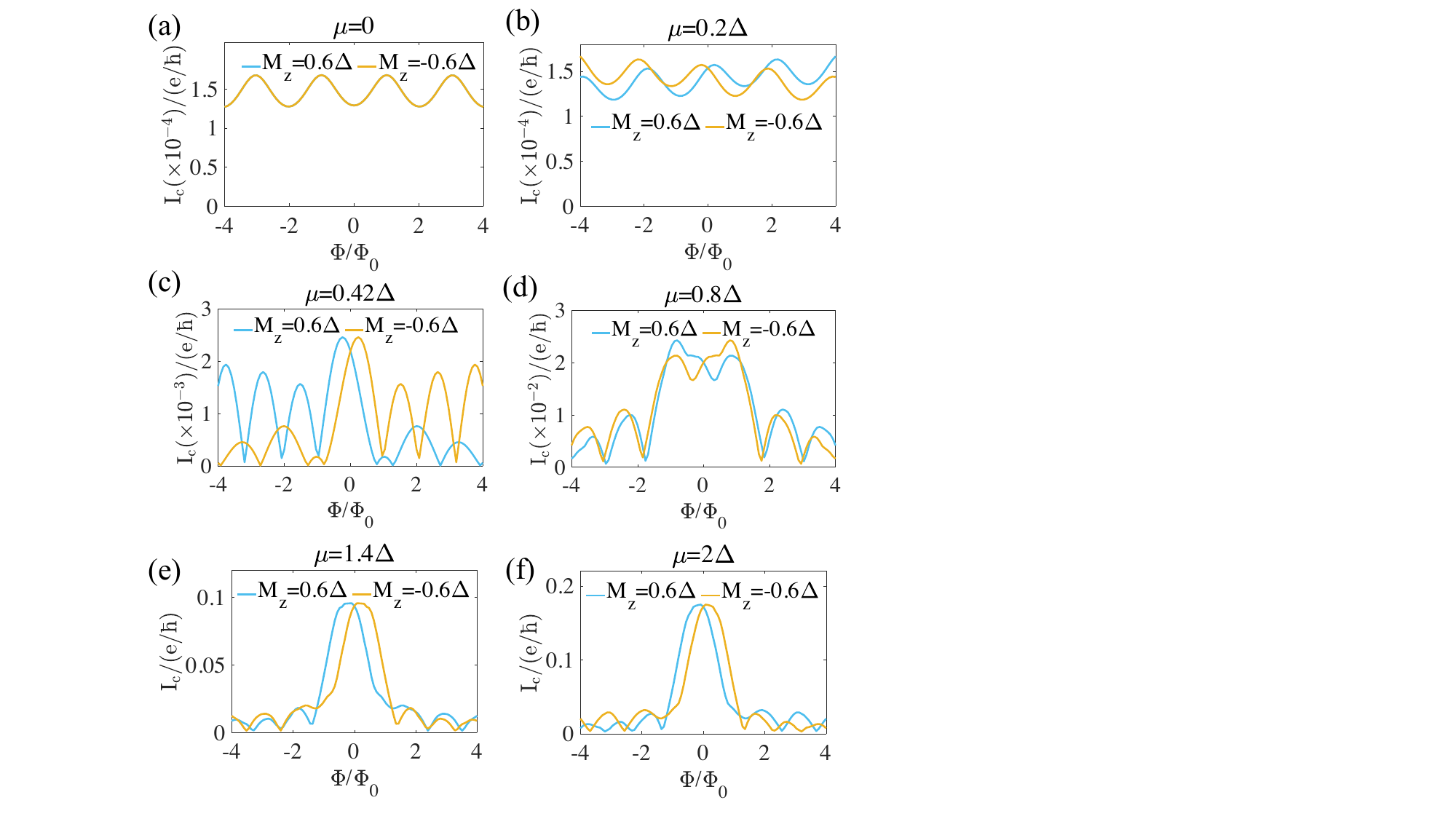}
\caption{Quantum interference patterns in different regimes. The critical Josephson current $I_c$ versus the magnetic flux $\Phi$ in the unit of $\Phi_0$. (a) CARs happen when only chiral edge states are included with $\mu=0$. $I_c$ exhibits the AB oscillation characterized by a periodicity of $2\Phi_0$.  (b)-(e) The behavior of $I_c$ deviates from the $2\Phi_0$-periodic oscillation as $\mu$ takes on values of $0.2\Delta$, $0.42\Delta$, $0.8\Delta$, and $ 1.4\Delta$,  respectively.  (c)  LARs take place in the JJs, and the $2\Phi_0$-periodic oscillation is destroyed. (d) The emergence of central  and side lobes is a consequence of the interference pattern transitioning towards a regime dominated by bulk carriers. 
(f) $I_c$ displays the typical asymmetric Fraunhofer pattern with $\mu=2\Delta$. In this regime, the bulk carriers uniformly flow through the whole QAHI region. In all cases,  the blue (orange) solid lines represent $I_c$ curves corresponding to  $M_z=\pm 0.6\Delta$, $\mu_s=\mu$, and the temperature $T=\Delta/200$ .}
\label{fig:2}
\end{figure}

\begin{figure}[ht!]
\centering
\includegraphics[width=1\linewidth]{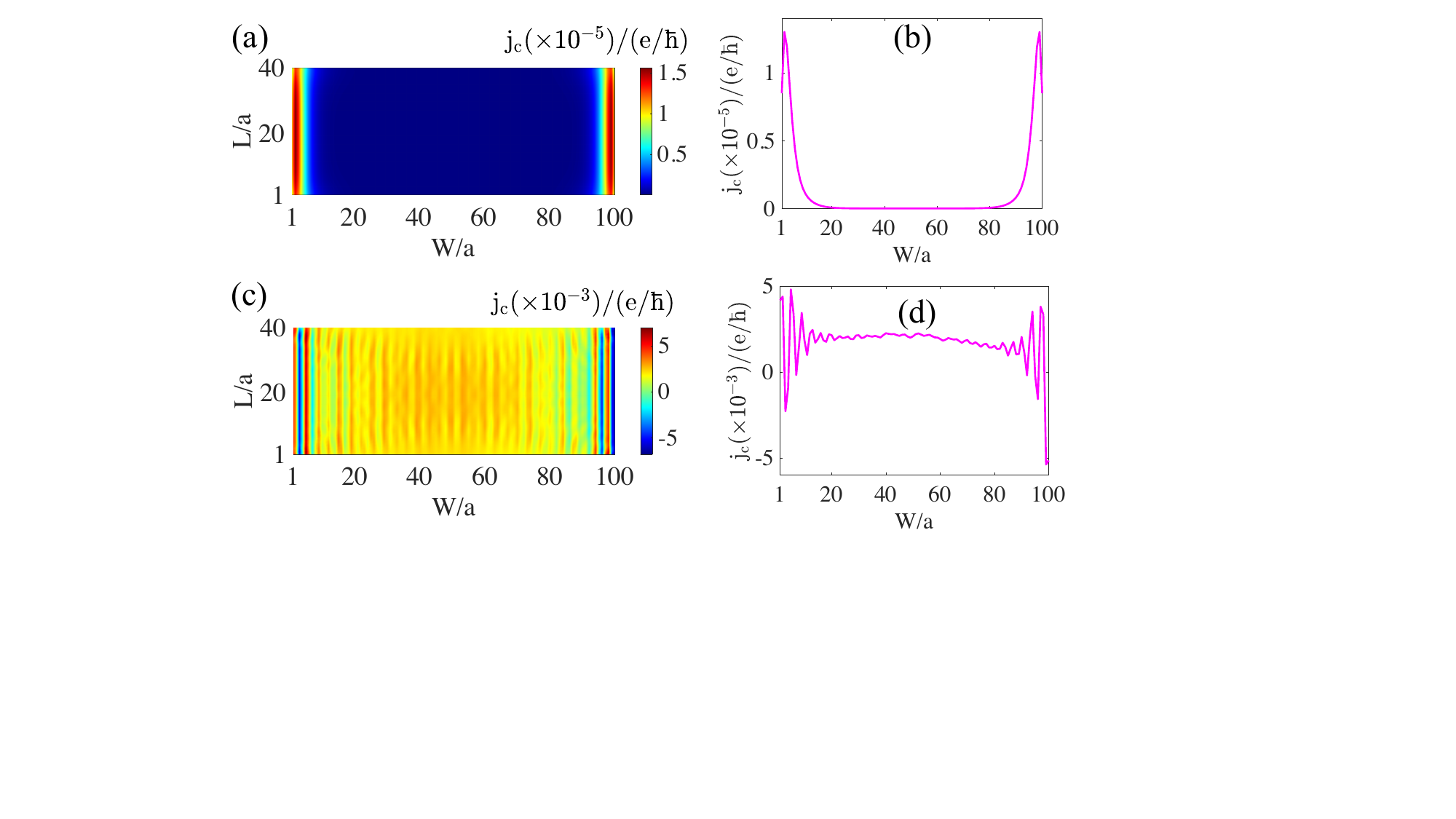}
\caption{The spatial distribution of current in the QAHI region [(a) and (c)] and at the QAHI-SC interface [(b) and (d)]  when the total current has its maximum value is examined under varying chemical potentials: $\mu=$0 [(a),(b)], and $2\Delta$ [(c),(d)]. In all cases, $M_z=0.6\Delta$, $\Phi=0$, $\mu_s=\mu$, and the temperature $T=\Delta/200$.  }
\label{fig:3}
\end{figure}

\section{Numerical results}\label{sec:results}

\subsection{ Evolution of quantum interference patterns}\label{subsec:B}

 In this section, we investigate the evolution of the quantum interference patterns within the QAHI-based junction as the chemical potential varies.  At a chemical potential of $\mu=0$, as depicted in Fig.~\ref{fig:2}(a), the QAHI region exclusively accommodates chiral edge carriers. 
At the interface between the QAHI and the SC, an electron from one chiral edge mode can tunnel into a hole on the opposite edge through the CAR process. Similarly, the holes are then reflected as electrons along the other QAHI-SC interface. This entire process forms a loop, completing the transfer of the supercurrent, and leading to an intriguing AB oscillation pattern. This interference pattern, a hallmark of chiral edge states, exhibits two distinctive features: (1) Unlike the conventional $\Phi_0$ periodicity, the critical Josephson current $I_c$ has a periodicity of $2\Phi_0$. (2) The minima of $I_c$ are non-zero.  These findings align with prior research studies \cite{f1,f2,f3,f4,f5,f6,f7,f8}. 
Furthermore, by shifting the chemical potential to $\mu=0.2\Delta$, an observed phase shift arises from the finite chemical potential. This occurrence is tied to the sustenance of the Josephson current by Andreev bound states forming a closed loop. Consequently, the product of the Fermi wavelength and the circumference, $k_F * L_d$, introduces a phase shift, resulting in current oscillations relative to the chemical potential.  {\color{black} The results depicted in Figs.~\ref{fig:2}(a) and (b) demonstrate that non-vanishing supercurrent can be sustained by the chiral edge states even in the absence of bulk carriers, exhibiting a periodicity of 2$\Phi_0$. Such a supercurrent feature of the chiral edge state is very distinct from the behaviors of bulk states in the conventional JJs.}
Interestingly, when $\mu$ to $0.42\Delta$, the chemical potential touches with the bottom of the conduction band (see Fig.~\ref{fig:1}(b)). Thus, bulk carriers begin to participate in the transport process, leading to the disruption of the $2\Phi_0$-periodic oscillation.

To trace the evolution of the Josephson current from edge states to the bulk further, we continue to increase the chemical potential to $\mu>0.42\Delta$, disrupting the previously observed $2\Phi_0$ periodicity.
Interestingly, the Fraunhofer pattern with a distinct central peak of width $2\Phi_0$ forms when $\mu>0.8\Delta$ (Figs.~\ref{fig:2}(d)-(f)), signaling the increased dominance of bulk states in the transport process. This occurrence correlates with the population of bulk carriers within the QAHI region, leading to a consistent appearance of LARs at QAHI-SC interfaces as $\mu$ increases (refer to Fig. \ref{fig:1}(d)). 
Due to the increase in bulk carriers, a noticeable enhancement in the maxima of $I_c$ is also observed, as seen in Figs.~\ref{fig:2}(a)-(f), exceeding hundreds of times.
Furthermore, the Fraunhofer pattern  displays an asymmetric feature due to broken time-reversal symmetry, namely, $I_c(M_z,\Phi) \neq I_c(M_z,-\Phi)$ when the magnetic field direction is reversed.  
Besides, the inversion symmetry invariance of {\color{black}the QAHI's Hamiltonian $H(\mathbf{k})$ } leads to the relationship $I_c(M_z,\Phi)=I_c(-M_z,-\Phi)$ \cite{S1,S2}. {\color{black}The inversion symmetry holds because $\mathcal{P} H(-\mathbf{k}) \mathcal{P}^{-1} = H(\mathbf{k})$ with $\mathcal{P}=\sigma_z \tau_0$, where $\sigma_z$ and $\tau_0$ are the Pauli matrix and the unit matrix in spin space and layer space, respectively. }
In the subsequent calculations, we focus exclusively on the scenario where $M_z=0.3$ due to the aforementioned relationship {\color{black} between $I_c(M_z)$ and $I_c(-M_z)$}. 
In summary, $I_c$ of the QAHI-based JJs  undergoes a
transition from a $2\Phi_0$-periodic oscillation pattern to an asymmetric
Fraunhofer pattern by tuning the chemical potential $\mu$.
 Note that we do not discuss the case where the chemical potentials $\mu$ and $\mu_s$ are different here because the periodicity of $2\Phi_0$ remains consistent when the difference is not large (See Appendix A for more information).

\begin{figure}[ht!]
\centering
\includegraphics[width=1\linewidth]{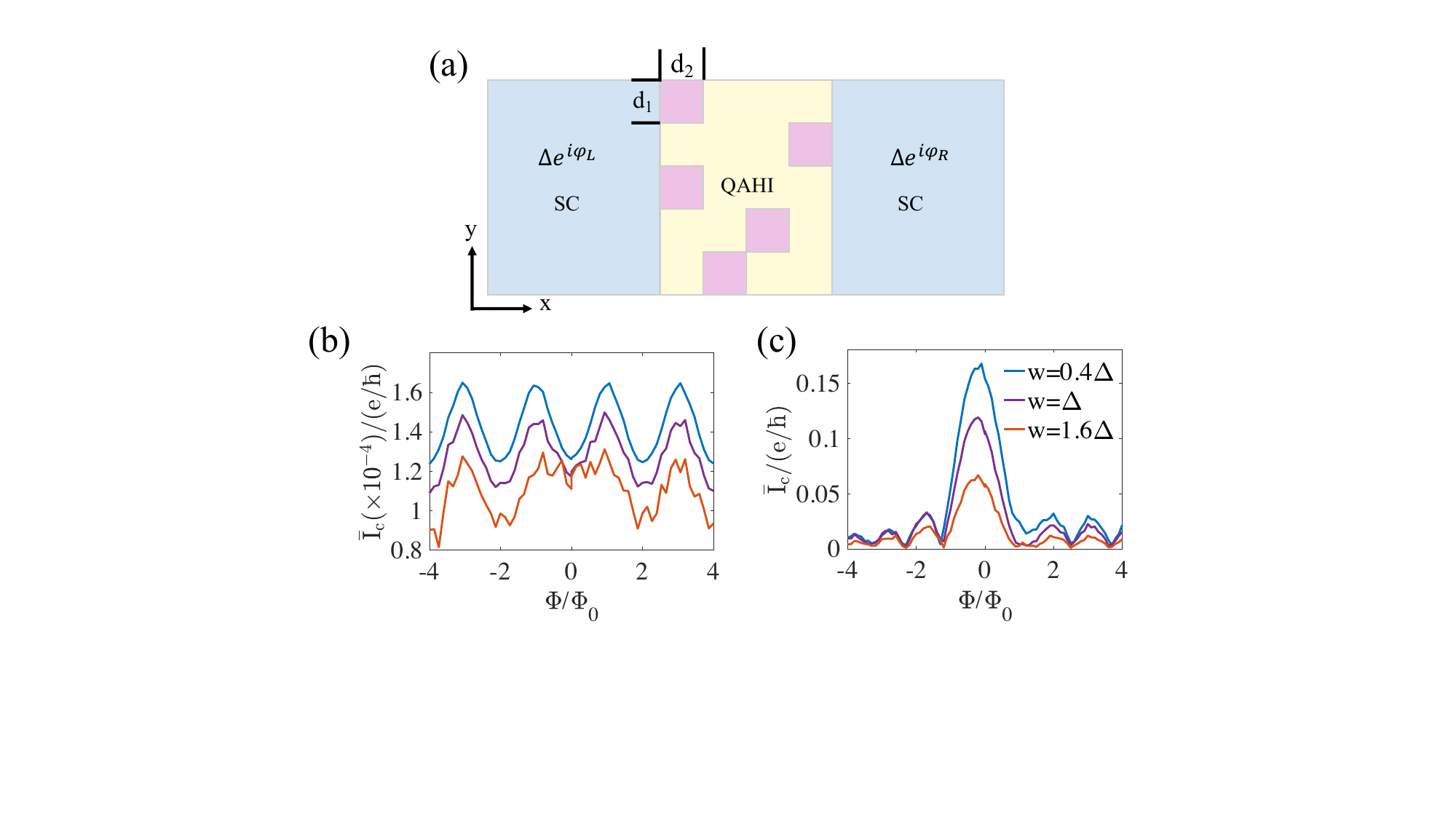}
\caption{(a) Schematic diagram of a QAHI-based JJ with a random {\color{black}magnetic domain} structure (top view). The {\color{black} domain} size is denoted by $d_1 \times d_2$. The quantum interference patterns are affected by the different strengths of magnetic disorder $w$ when (b) $\mu=0$ and (c) $\mu=2\Delta$. In all cases, the {\color{black} domain} size is $d_1 \times d_2= 10a \times 4a$, and the {\color{black} domain} probability $P=0.5$.  The remaining parameters are set as: $M_z=0.6\Delta$, $\mu_s=\mu$, and the temperature $T=\Delta/200$. We average over 50 random {\color{black} domain} configurations \cite{Ave}.  }
\label{fig:4}
\end{figure}

To further reveal the origin of different interference patterns,
we investigate the distribution of current $j_c$ in QAHI region when the total current has its maximum value \cite{M1,M2,M3}. The local critical current is usually used to represent the current distribution of various systems, e.g., quantum Hall systems 
 \cite{LHL}. Figure~\ref{fig:3} demonstrates the  distribution of current along the x-direction, $j_c(y)$, corresponding to the superconducting phase difference $\varphi$, precisely when the total current $\sum\limits_{y=1}^{W} j_c(y)$ reaches its maximum.
At $\mu=0$, the current predominantly flows along the QAHI region's two edges, as evidenced in the spatial distribution of $j_c$ displayed in Figs.~\ref{fig:3}(a) and (b). This observation aligns with the phenomenon depicted in Fig.~\ref{fig:1}(c), where transport within the central region is solely governed by the propagation of chiral edge carriers. Additionally, this  distribution of current is consistent with the quantum interference pattern shown in Fig.~\ref{fig:2}(a), characterized by a periodicity of $2\Phi_0$. Moreover, Figs.~\ref{fig:3}(c) and (d) show a bulk-dominated transport process, as the $\mu$ value is increased to $2\Delta$. Thus, local Andreev pairs become uniformly distributed across the QAHI region (see Fig.~\ref{fig:1}(d)), resulting in a Fraunhofer pattern, as observed in Fig.~\ref{fig:2}(f).

\subsection{Effects of {\color{black}magnetic domains}}\label{subsec:C}

Given the inevitable appearance of {\color{black} domains} in magnetically doped topological insulators, we utilize a percolation model to simulate the effects of random {\color{black} domains} on quantum interference patterns \cite{Li}. 
In realistic samples, {\color{black} domains} typically exhibit irregular shapes. However, the specific shapes of {\color{black} domains} are not of particular importance to our study. The bulk carriers induced by {\color{black} domains} that are essential to the physics discussed in this section.  Therefore, for the sake of simplicity, we model {\color{black} domains} as rectangular in shape.

\begin{figure}[ht!]
\centering
\includegraphics[width=1\linewidth]{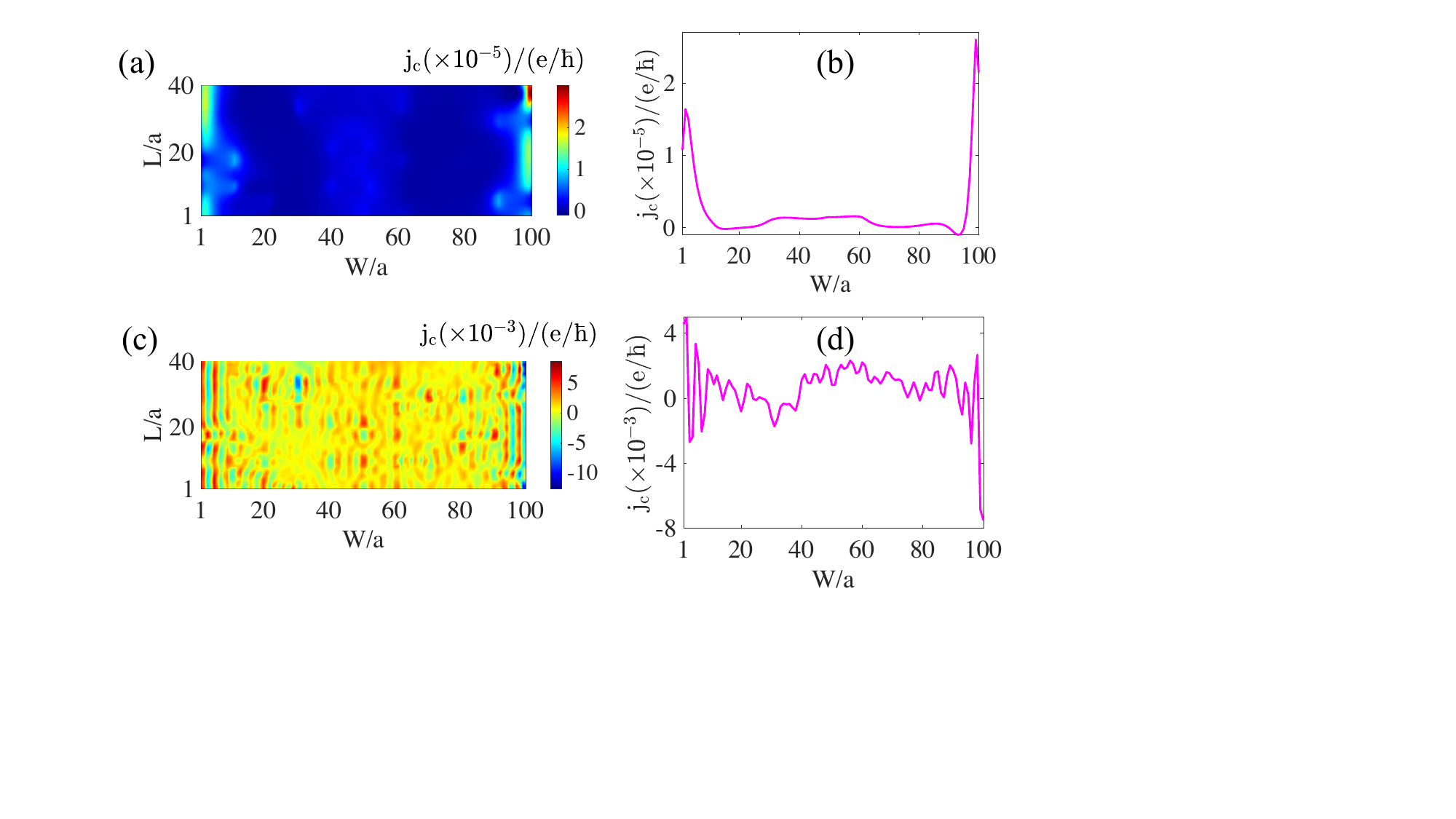}
\caption{The spatial distribution of current in the QAHI region [(a) and (c)] and at the QAHI-SC interface [(b) and (d)]  when the total current has its maximum value affected by the {\color{black} domain} structure is examined under varying chemical potential values: $\mu$=0 [(a),(b)] and $2\Delta$ [(c),(d)].  In this case, the {\color{black} domain} size is $d_1 \times d_2= 10a \times 4a$, the magnetic disorder strength $w=1.6\Delta$, and the {\color{black} domain} probability $P=0.5$.  In all cases, $M_z=0.6\Delta$, $\Phi=0$, $\mu_s=\mu$, and the temperature $T=\Delta/200$. }
\label{fig:5}
\end{figure}

Fig.~\ref{fig:4}(a) demonstrates the division of the central region into multiple blocks, each consisting of specific unit cells sized $d_1\times d_2$. 
The magnetization within each block is randomly chosen to potentially flip with a probability of $P$, while the flipped magnetization strength is uniformly distributed in the range of $[-w/2, w/2]$ with the disorder strength $w$.

In Figs.~\ref{fig:4}(b) and (c), we display the {\color{black} domain}-averaged critical Josephson current, denoted as $\overline{I}_c$. Initially, focusing on the case with $\mu=0$ in Fig.~\ref{fig:4}(b), it is observed that $\overline{I}_c$ remains relatively stable (in comparison with Fig.~\ref{fig:2}(a)) when $w=0.4\Delta$, indicating the robustness of chiral edge states against weak magnetic disorder ($w$). 
Then, at $w=\Delta$, the partial destruction of chiral edge modes diminishes the peak of the critical Josephson current $\overline{I}_c$. Despite {\color{black} domains} causing the emergence of some bulk carriers in the QAHI region, the $2\Phi_0$ periodicity remains notably stable, owing to the dominance of transport by chiral edge modes.  
After that, increasing $w$ to $1.6\Delta$ initiates two simultaneous processes: the breakdown of chiral edge modes and a comparable contribution of bulk carriers to that of chiral edge carriers. Consequently, both the $2\Phi_0$ periodicity and the characteristic shape of the quantum interference pattern are lost. 
Conversely, at $\mu=2\Delta$, we observe remarkable stability in the interference pattern of bulk carriers, maintaining the shape of the asymmetric Fraunhofer pattern even at $w=1.6\Delta$. This suggests that the unconventional $2\Phi_0$-periodic oscillation is more susceptible to disruption compared to the resilient nature of the asymmetric Fraunhofer pattern under {\color{black} domain} structures.
Furthermore, the values of $\overline{I}_c$ gradually decrease with increasing $w$ in Figs.~\ref{fig:4}(b) and (c), manifesting the influence of disorder effects.

\begin{figure}[ht!]
\centering
\includegraphics[width=1\linewidth]{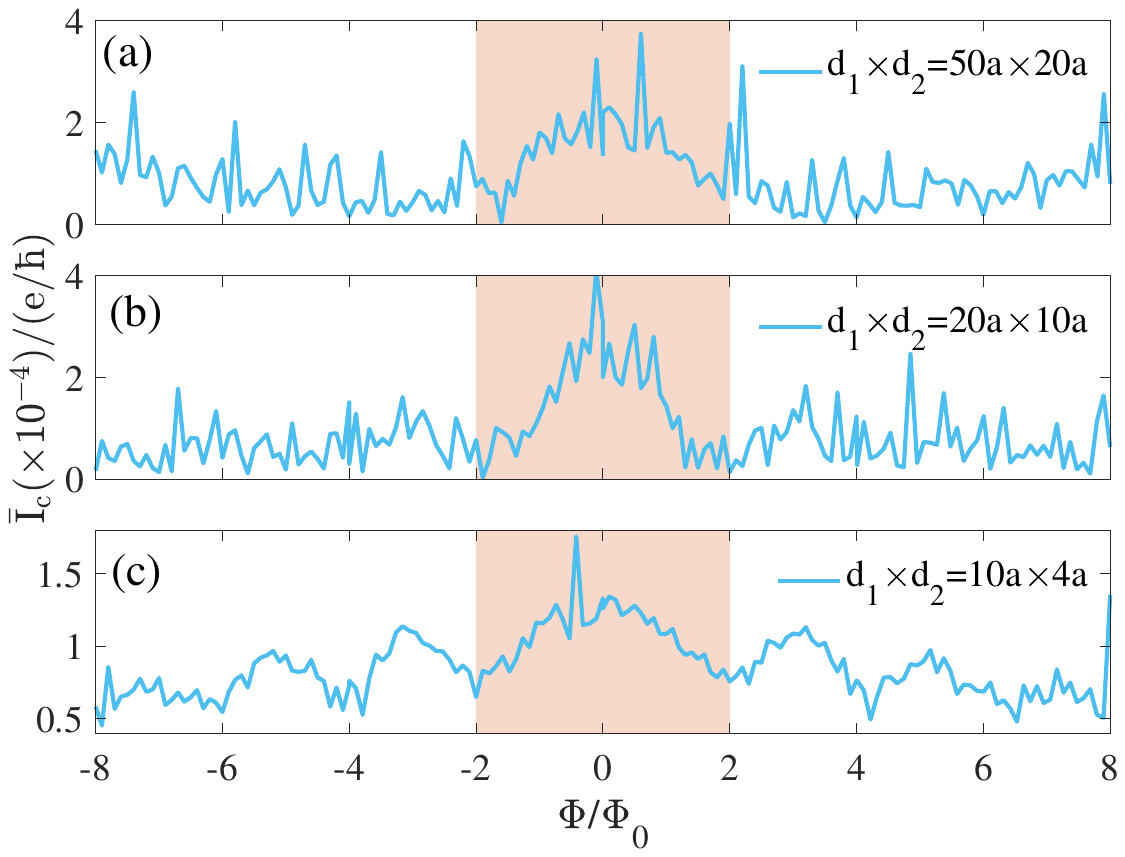}
\caption{The Fraunhofer-like patterns. The quantum interference patterns are affected by the {\color{black} domain} structure at $\mu=0$ in a larger magnetic field range with different block sizes $d_1 \times d_2$: (a) $50a \times 20a$, (b) $20a \times 10a$, and (c) $10a \times 4a$. In all cases, the magnetic disorder strength $w=2\Delta$ and the {\color{black} domain} probability $P=0.5$.  The remaining parameters are set as: $M_z=0.6\Delta$, $\mu_s=\mu$, and the temperature $T=\Delta/200$. In (a)-(c), the orange regions show the central lobes of the Fraunhofer-like pattern. We average over 15,25,50 random {\color{black} domain} configurations in (a),(b), and (c), respectively. }
\label{fig:6}
\end{figure}

Next, we examine  the spatial distribution of current $j_c$ within the QAHI region and at the QAHI-SC interface  when the total current has its maximum value in Fig.~\ref{fig:5}, considering the existence of {\color{black} domain} structures. 
In our calculation, we maintain a constant total magnetic flux $\Phi=0$ and compute the  distribution of current $j_c(y)$ along the x-direction when the total current $\sum\limits_{y=1}^{W} j_c(y)$ reaches its maximum.
Due to the presence of {\color{black} domain} structures, the  distribution of current $j_c$ exhibits static eddy-like currents in addition to constant currents. In Figs.~\ref{fig:5}(a) and (b) at $\mu=0$, it is evident that the chiral edge modes are disrupted, and some bulk carriers emerge within the inner QAHI region due to the random distribution of {\color{black} domains} across the QAHI region. This observation aligns with the behavior depicted in the curve corresponding to $w=1.6\Delta$ in Fig.~\ref{fig:4}(b).
Conversely, when $\mu=2\Delta$, the distribution of $j_c$ exhibits characteristics akin to typical bulk transport, facilitating the stability of the asymmetric Fraunhofer pattern within {\color{black} domain} structures. Notably, in Figs.~\ref{fig:5}(c) and (d), $j_c$ values oscillate more prominently compared to Figs.~\ref{fig:3}(c) and (d).

To gain deeper insights into how {\color{black} domains} affect the edge current, we explore the quantum interference pattern at $\mu=0$ by varying the sizes of {\color{black} domains} under a strong magnetic disorder strength $w=2\Delta$, spanning a wider range of magnetic fields.
In Fig.~\ref{fig:6}(a), an emergent central lobe is observed within the interval $\Phi/\Phi_0 \in (-2, 2)$, alongside side lobes at higher values of $\Phi$. 
Additionally, the increase in the maxima of the central lobe is linked to the emergence of bulk carriers caused by a limited number of {\color{black} domains}, particularly when the {\color{black} domain} size is $d_1 \times d_2 = 50a \times 20a$.
This interference pattern remarkably resembles  to the Fraunhofer pattern illustrated in Fig.~\ref{fig:2}(f). The emergence of bulk carriers in the QAHI region due to the presence of {\color{black} domains} is the underlying cause of this phenomenon. 
When we decrease the {\color{black} domain} size to $d_1 \times d_2 = 20a \times 10a$ to allow for a larger quantity of {\color{black} domains}, a more distinct Fraunhofer-like pattern is evident within the $\overline{I}_c$ values in Fig.~\ref{fig:6}(b). 
Again, the presence of bulk carriers contributes to the enhancement of maxima of the central lobe, compared to the scenario depicted in Fig.~\ref{fig:2}(a).
When the {\color{black} domain} size is further decreased to $d_1 \times d_2 = 10a \times 4a$, as depicted in Fig.~\ref{fig:6}(c), the increased visibility of the periodicity in the higher lobes reveals a distinct Fraunhofer-like pattern. In particular, the periods of the central and side lobes are twice those of the conventional Fraunhofer pattern.
To conclude, the observed Fraunhofer-like pattern in chiral JJs likely extends beyond our model and is prevalent in those doped with magnetic impurities, holding promise for further experimental verification. In summary, the interference pattern within the {\color{black} domain} structure can be influenced by factors such as magnetic disorder strength $w$ and {\color{black} domain} size $d_1 \times d_2$.

\section{discussion and conclusion}\label{sec:conclusion}

{\color{black} While QAHIs can manifest in various material classes, our work focuses exclusively on magnetically doped topological insulators. The lack of comprehensive experimental results on QAHI-based JJs is primarily due to the significant experimental challenges. This challenge arises from the difficulty of achieving coexistence among the QAHI and superconductivity. Specifically, realizing the QAHIs requires magnetic doping within topological insulators. Although magnetic dopants are crucial for realizing QAHI, they also pose challenges for maintaining superconductivity. Therefore, achieving an appropriate magnetic doping ratio is crucial in experimental setups. Additionally, the presence of magnetic dopants raises questions regarding the formation of domains, which is one of the primary aspects addressed in our work. This results in the appear of the anomalous Fraunhofer-like pattern, which is  likely widespread in chiral JJs doped with magnetic impurities.  }

In this work, we explore the transport properties of the chiral JJs based on the QAHIs connecting to two superconducting electrodes, through numerical calculation of the Josephson current ($I_c$). We present a systematic transition from edge-state to bulk-state dominated supercurrent  as the chemical potential varies. This transition leads to an evolution from a $2\Phi_0$-periodic oscillation pattern to an asymmetric Fraunhofer pattern. Furthermore, we observe the emergence of a novel Fraunhofer-like pattern, characterized by periods twice those of the conventional pattern. This emergence is attributed to the coexistence of chiral edge states and bulk states induced by {\color{black}magneitc domains}, even when the chemical potential is within the gap.   This phenomenon is likely widespread in chiral JJs doped with magnetic impurities. Our findings also hold promise for direct experimental verification.

\begin{acknowledgments}

The authors would like to thank Yang Feng, Yu-Hang Li, and Qing Yan for helpful discussions. We are grateful to the support by the National Key R\&D Program of China (Grants No. 2022YFA1403700),
the National Natural Science Foundation of China (Grant No.12204053), and the Innovation Program for Quantum Science and Technology (Grant No.2021ZD0302400).
C.-Z. Chen is also support by the Natural Science Foundation of Jiangsu Province Grant (No. BK20230066)
and the Priority Academic Program Development (PAPD) of Jiangsu Higher Education Institution.
\end{acknowledgments}

\begin{appendix}

\section{The effects of the mismatch between $\mu$ and $\mu_s$}\label{sec:A}

\begin{figure}[h]
    \centering
    \includegraphics[width=1\linewidth]{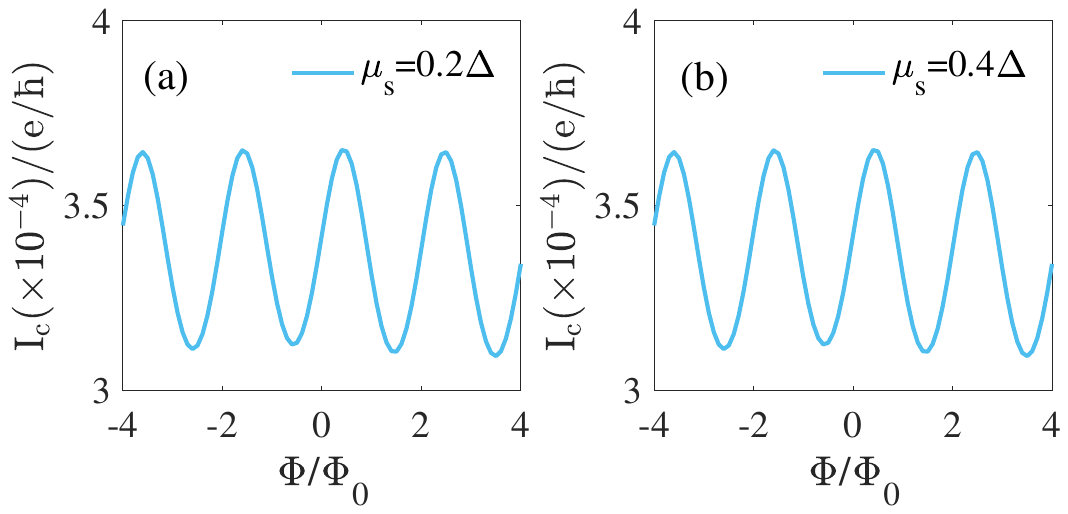}
    \caption{ Quantum interference patterns when the chemical potentials $\mu$ and $\mu_s$ exhibit differences. The critical Josephson current $I_c$ versus the magnetic flux $\Phi$ in the unit of $\Phi_0$ with varied chemical potentials in the superconducting regions:  (a) $\mu_s=0.2\Delta$, and (b) $\mu_s=0.4\Delta$. In all subplots, we take $\mu=0$, and the remaining parameters are consistent with those in Fig. 3(a) in the main text.}
    \label{fig:7}
\end{figure}

We investigate the effects of the mismatch between $\mu$ and $\mu_s$ on the quantum oscillation patterns. It should be pointed out that the mismatch in chemical potentials alters the scattering probability of Andreev reflections, potentially resulting in three aspects: (1) deviations from the periodicity of $2\Phi_0$, (2) the maxima of the critical current occurring not at $\Phi=\pm(2n+1)\Phi_0$, where $n$ is an integer, and (3) variations in the values of the critical currents. However, it should be emphasized that when the difference between $\mu$ and $\mu_s$ is not large, the periodicity of $2\Phi_0$ remains very steady, as illustrated in Fig.~\ref{fig:7}. Consequently, we set $\mu_s=\mu$ in the main text.

\end{appendix}

\bibliographystyle{apsrev4-1}

\begin{thebibliography}{100}
\bibitem{Ave}
  We don't conduct averaging over more samples. The main reason for this limitation is the significant challenge posed by the computational constraints in numerical simulations. Furthermore, the fluctuations observed in Figs. 5(b), 6 and 8 result from insufficient ensemble averaging.
\end{thebibliography}

\end{document}